\documentclass[oldversion]{aa}
\usepackage{graphicx}

\begin{document}

\title{Testing the cosmic ray content in galaxy clusters}

   \author{P. Marchegiani \inst{1}, G.C. Perola \inst{1}, and S. Colafrancesco \inst{2}}

   \offprints{G. C. Perola}

\institute{   Dipartimento di Fisica, Universit\`a Roma 3, Via della Vasca Navale 84, Roma, Italy
              Email: marchegiani@fis.uniroma3.it, perola@fis.uniroma3.it
\and
              INAF - Osservatorio Astronomico di Roma
              via Frascati 33, I-00040 Monteporzio, Italy.
              }

\date{Received / Accepted }

\authorrunning {P. Marchegiani et al.}

\titlerunning {Cosmic rays in galaxy clusters}

\abstract{The effective content of cosmic rays ($CR$) in galaxy clusters remains elusive. The evidence
of relativistic electrons ($RE$) in the subset of clusters endowed with a radio halo
remains  hardly
quantitative in the absence of robust estimates of the magnetic field $B(r)$, derived from
Faraday Rotation (FR) measurements. The content in relativistic protons ($RP$) requires a different
approach, the only direct one residing in the detection of their collisional production of
gamma rays ($GR$).

Based on the evidence of merging phenomena in clusters, theory predicts a large content
of $RP$, whose energy density could be a large fraction of the thermal energy. This paper aims to
estimate a maximum production of both secondary relativistic electrons, $SRE$, and $GR$ from the
$RP$ that have supposedly accumulated throughout the entire history of a cluster.

$SRE$ and $GR$ production is maximized when the $RP$ and the thermal gas share the
same radial profile. The production rate is normalized by adopting a reference value of 0.3 for $\xi$, the ratio of $RP$
to thermal pressure.  The $SRE$ content which obtains, when constrained to reproduce
the observed radio brightness profile, yields univocally $B(r)$, if the presence of primary $RE$ were negligible.

This procedure is applied to four radio--halo clusters (Coma, A2163, A2255, A2319).
In these objects, the central value $B_0$ required is consistent with typical,
albeit rather uncertain, values derived from FR, 
although for A2163 and A2319 no reliable FR estimates are available to strengthen this result.
On the other hand, $B(r)$ typically increases beyond the thermal core, a hardly acceptable condition.
This problem is alleviated by assuming
a mix of $SRE$ and of ``primary'' $RE$ ($PRE$), with the latter
becoming the dominant component beyond the thermal core. These results suggest that in clusters without
a radio halo detected so far
a diffuse radio-emission should also be observable due to $SRE$ alone, and therefore more
centrally condensed, provided that $\xi$ is of the order of 0.3. 
To encourage deeper radio observations of such clusters, some examples
were selected that seem rather promising. Efforts in this direction, if accompanied by FR measurements,
could provide highly significant constraints on the $CR$ content in clusters, even before
the future GLAST mission will have accomplished the hard task of detecting the $GR$.
A complementary result concerns the excess far UV in the Coma cluster, that some authors
have attributed to IC emission from $SRE$. It is shown that this hypothesis can be excluded,
because it requires a $RP$ energy content in excess of the thermal one.}

\keywords{Cosmology; Galaxies: clusters; cosmic-ray: origin}

 \maketitle


\section{Introduction}

A significant fraction of galaxy clusters are endowed with a
diffuse, low-brightness radio-emission, which is clearly distinct from the
extended radio components, if present, associated with member
galaxies. The historical prototype (Willson 1970) is the radio source Coma C
in the Coma cluster. These so-called radio halos represent
direct evidence of relativistic electrons and magnetic fields
embedded in the thermal plasma responsible for the X-ray emission.
It is very likely that relativistic protons are present as
well. In the following these components will be labelled
$RE$ (electrons), $RP$ (protons), $CR$ (for cosmic rays, protons and electrons),
and $B$ (the magnetic field and its strength). Since the brightness of the
radio emission depends (non-linearly) on $B$ and the $RE$ density,
it cannot be excluded that the presence of a non-thermal component is
common to all clusters, including those without, as yet, an
observed radio-halo.

Generally speaking, we are still far from solid quantitative
estimates of the non-thermal content. The radio emission alone
can only provide a minimum for the energy content in $CR$ and $B$
combined, according to the recipe first proposed by Burbidge (1956).
This minimum is well below the thermal energy content (Govoni
\& Feretti 2004).

A direct way to estimate the $RE$ and $B$ contents separately was
first applied by Perola \& Reinhardt (1972) to Coma C. Although at
the time they could (incorrectly, as it became clear later) and
did attribute the whole of the X-ray emission to this process, the
$RE$ inevitably lose energy on the photons of the cosmic
microwave background through the so-called inverse Compton (IC)
process, thus it remains obviously true that, within a rather wide
range of $B$ values, the energy lost should be observed in the
X-ray band. All the attempts, made so far, to disentangle this contribution from
the thermal X-ray emission, which have concentrated attention
on the hard (greater than $10-30$ keV) regime, have led to
results that are rather controversial particularly in their
interpretation (Fusco-Femiano et al. 2004; Rossetti \& Molendi
2004). This follows from the evidence of an excess,
possibly IC emission, obtained with non-imaging
instruments, which therefore can only be used to estimate the
volume-integrated amount of $RE$ within the instrument's field of
view and, rigorously speaking, only in a very narrow energy
interval, $1.6-3.2$  GeV. In a previous paper (where most
references can be found, Colafrancesco, Marchegiani \& Perola
2005, Pap. I), we discussed this issue at length and
emphasized the need to measure the surface brightness
distribution of the IC emission, in order, first, to verify whether the $RE$
responsible for this radiation are effectively confined within the
observed radio halo and, second, to solve the degeneracy intrinsic to
the synchrotron emission and therefore obtain fundamental
information on the radial dependence of the $RE$ density and of
$B$.

The origin of the $RE$ was outside the goal of Paper I. They can have
a ``primary'' origin, which is accelerated along with the $RP$.
However, there must also be $RE$ of ``secondary'' origin, produced
through the inevitable inelastic interactions of the $RP$ with the
ambient plasma. This process, notably, gives rise to the
emission of gamma rays ($GR$) with energy $\geq$ 50 MeV. If we
knew the number and distribution of the $RE$ (from their IC
emission), the question (Dennison 1980) whether the ``secondary''
ones ($SRE$) are either a negligible or an important fraction of
the whole $RE$ population would require assumptions on the amount
of $RP$, with basically one single important constraint, namely
that their pressure should not exceed that of the thermal gas.
Vice versa, if by measuring the spectrum, flux, and
the associated surface brightness of the gamma rays (after
verifying that they cannot be due to other mechanisms, such as the
hypothetical annihilation process of the dark matter, see for
instance Pieri \& Branchini 2004, Colafrancesco et al. 2006) one
could derive the amount of $RP$, a very stringent constraint on
the production rate of the $SRE$ could be placed. If this rate
were too low for a significant contribution to the radio-emission,
except for unacceptably high values of $B$, the answer to the
question would be immediate. It is clear, however, that
only the combination of reliable measurements of the $GR$ and the
IC X-rays, with the latter leading to an absolute estimate of the $RE$
amount and distribution, will close the circle. The amount and
distribution of the $RE$ could also be obtained via an
independent estimate of $B$ (radial distribution included), which
can be derived from Faraday rotation (FR) measurements, once this
goal is reached in an uncontroversial way (see Govoni \&
Feretti 2004 for a discussion of the state of this art). In either
(IC or FR) case, the $GR$ represent an irreplaceable step for the
issue of the $CR$ content.

In this context, the contribution that the present paper
aims to provide, given the still immature observational status,
consists in evaluating the ``maximal'' requirements on the
$RP$ and on $B$, including their radial distribution, assuming
that the radio-emission is ``totally'' due to the $SRE$. These
requirements, constrained to avoid conflicts with either physical arguments
(the non-thermal pressure should not exceed the thermal pressure)
or observational evidence (limits on $B$ from FR), will be used
to predict the uppermost value
of the $GR$ emission intensity, which one would expect to inevitably follow.

The authors are aware of the fact that arguments have been put
forward that are against a dominant contribution to the $RE$ of
the $SRE$ component (e. g. Brunetti 2003). Nonetheless, the
approach adopted is useful on two different grounds. First, the
$RE$ lifetime in the energy range of interest is short (about one
tenth) compared to the age of a cluster; thus the currently
relevant primary $RE$ ($PRE$) must have been accelerated in this
last fraction of the cluster age, while the $RP$ giving rise to
the $SRE$ in the same fraction of time have likely accumulated over
the entire cluster lifetime, meaning that their contribution to the $RE$
content might be far from negligible with respect to that due to
the acceleration processes. Second, the choice of maximizing this
contribution leads in a sense to the most optimistic predictions
on the intensity of the $GR$ emission, which could be immediately
tested by its measurement.

Another item is connected with the excess emission reported
in the far ultraviolet-soft X-rays in the Coma cluster in
particular, that might be attributed to IC emission. If
this were the case, the relevant $RE$ population would fall into a totally
different range of energies with respect to the radio-emitting
ones. In this hypotheses, some authors (Bowyer et al. 2004) have attributed
a ``secondary'' origin to these $RE$. Its relevance to the
ensuing requirements on the parent $RP$ will also be discussed.

The paper is organized as follows. Section 2 contains a
brief overview of the theoretical results on the acceleration
of $RP$ in clusters. Section 3 is devoted to the assumptions and prescriptions
on which
the subsequent calculations are based. Section 4 illustrates the properties of
the four clusters used as examples and the way their radio brightness
distribution is described analytically. In Section 5 the results are reported
for the $GR$ emission and for the strength and radial distribution of $B$,
when the radio halo is ``totally'' attributed to the $SRE$.
Section 6 is devoted to the issue of the UV excess in the Coma cluster
as due to IC from $SRE$. Section 7 is devoted to a discussion, and Section 8
to the conclusions.

\section{Acceleration and accumulation of $RP$}

Since Jaffe (1977) discussed the possible diffuse origin
(that is, {\it in situ} acceleration {\it versus} injection by radio sources)
of the $RE$ in Coma C, science has progressed a long way.
In the theoretical scenarios of structure formation in the
Universe through hierarchical clustering, with the relatively
recent support from Chandra and XMM-Newton observations, the
merging processes are widely suspected to be the events that
can lead to a diffuse acceleration of $CR$ in clusters, along with an increase
in the energy content of the thermal gas.

The papers quoted below do refer to the results (Blandford \& Eichler 1987)
of the shock diffusive acceleration
theory that predict spectral indexes $s$ linked to the Mach number $M$ as

\begin{equation}
s=2\frac{M^2+1}{M^2-1}.
\label{mach}
\end{equation}

From hydrodynamical simulations (e. g. Ryu et al. 2003) it appears
that in the central regions, where the largest amount of particles
are processed, the typical values of the Mach number are 
$2\leq M \leq 4$, and, according to Eq. (\ref{mach}), 
those of spectral index are $2.3
\leq s \leq 3.3$. The amount of energy going into the $CR$ channel
can be as large as $30-50\%$ of what goes into the thermal
channel (Miniati et al. 2001a; Ryu et al. 2003). Much higher
values of $M$, which are even more efficient in the acceleration
process, are attained in the cluster outskirts, where (Pfrommer et
al. 2006) a higher ratio between the two energy contents could
eventually prevail, except that dynamical stability problems might
then ensue. In this paper, as explained in the
next section, care is taken that nowhere within the
radio halos, as defined by the outermost contour of the
available maps, does the non-thermal pressure (inclusive of $CR$ and
$B$) exceed the thermal pressure. Furthermore, a
single value of $s$ is adopted for simplicity, namely $s=2.7$. (Incidentally,
this is the value required for reproducing with a $SRE$
population the radio slope of Coma C, the only one well determined
among the four cases which will be considered in the following).
A value of $s$ higher than 2
implies that the choice of the minimum energy is critical to the $CR$ pressure.
Following Gieseler, Jones \& Kang (2000), the minimum adopted Lorentz
factor is given by

\begin{equation}
\gamma_{min} \approx 1+ \left[3.4\times10^{-5}
kT_{keV} \left( \frac{c_2}{4} \right)^2\right],
\label{gammamin}
\end{equation}
where $kT_{keV}$ is the thermal gas temperature, and $c_2$ is
approximately equal to 4. Thus $\gamma_{min}$ is practically equal to 1
in all cases.

For the $RP$ momentum ($p$) spectrum, the form adopted is
$N_{RP}(p)\propto p^{-s}$, which can be written as a function of $\gamma$:

\begin{equation}
 N_{RP}(\gamma,r) = K_{RP}(r) \frac{\gamma}{(\gamma^2-1)^{(s+1)/2}}.
\label{eq.spettro.protoni}
\end{equation}
The associated pressure is given by

\begin{equation}
P_{RP}(r)=\frac{1}{3}m_pc \int_{\gamma_{min}}^{\gamma_{max}}
N_{RP}(\gamma,r) \beta \gamma v(\gamma) d\gamma.
\label{eq.press.cr}
\end{equation}
The radial dependence of the RP content, with a constant
$s$, as anticipated, will be described through
that of their $P_{RP}$.

\section{Assumptions and prescriptions}

The production rates per unit volume of the $SRE$, $Q_e(\gamma_e,r)$ and
of the $GR$, $Q_g(E_g,r)$, which are strictly related to each other,
are proportional to the product of the thermal and
non-thermal proton density. Since the first of the two quantities
is known, the radial behavior of the second, which maximizes the
production rates, must be found.

According to what has been described in Sect. 2, and assuming that among the $CR$
the $RP$ are the energetically dominant component, it is
convenient to introduce the ratio between $P_{RP}$ and $P_{th}$, the
pressures of the non thermal and the thermal components,

\begin{equation}
\xi(r)=\frac{P_{RP}(r)}{P_{th}(r)},
\label{xi}
\end{equation}
which is bound to remain everywhere less than unity:

\begin{equation}
\xi(r) \leq 1.
\label{xileqone}
\end{equation}

So long as $kT$ can be regarded as constant (a reasonably good description
of clusters within the extent of their radio halos), and under the assumption
that the mean $RP$ energy is also constant (a direct consequence of the
assumption of a constant value of the exponent $s$ anticipated
in Sect. 2), $\xi(r) \propto n_{RP}/n_{th}$,
where $n_{RP}$ ($\equiv \int N_{RP}d\gamma$) and $n_{th}$ are the number densities of the relativistic
and the thermal protons, respectively. Without loss of generality,
the radial dependence of $n_{RP}$ can be chosen with the same
functional form as is generally used for $n_{th}$,

\begin{equation}
n_{th}(r)=n_{th,0} \left[1+\left(\frac{r}{r_{c,th}}\right)^2\right]^{-q_{th}},
\label{radialthermal}
\end{equation}
with the two parameters, core radius and exponent, left free:

\begin{equation}
n_{RP}(r)=n_{RP,0} \left[1+\left(\frac{r}{r_{c,RP}}\right)^2\right]^{-q_{RP}}.
\label{radialrelprot}
\end{equation}

Not surprisingly, the production rates reach their maximum values
when $r_{c,RP}$ = $r_{c,th}$ and $q_{RP}$ = $q_{th}$, therefore
for $\xi$ constant with radius, which will be labelled $\xi_0$.
These values are given by

\begin{eqnarray}
 Q_e(\gamma_e,r) &= & n_{th,0}K_{RP,0}\left[ 1+ \left(\frac{r}{r_{c,th}}\right)^2
\right]^{-2q_{th}} \int_{\gamma_{min}}^\infty c \beta(\gamma_p) \nonumber \\
 & & \times
\frac{\gamma_p}{(\gamma_p^2-1)^{(s+1)/2}}
 \sigma_{\pi^\pm}(\gamma_p) F_e(\gamma_e,\gamma_p) d\gamma_p
\label{elprodrate}
\end{eqnarray}

 \begin{eqnarray}
 Q_g(E_g,r) &= & n_{th,0}K_{RP,0}\left[ 1+ \left(\frac{r}{r_{c,th}}\right)^2
\right]^{-2q_{th}} \int_{\gamma_{min}}^\infty c \beta(\gamma_p)
 \nonumber \\
 & & \times \frac{\gamma_p}{(\gamma_p^2-1)^{(s+1)/2}}
 \sigma_{\pi^0}(\gamma_p) F_g(E_g,\gamma_p) d\gamma_p
\label{gprodrate}
 \end{eqnarray}
where $\sigma_{\pi^\pm}$ and $\sigma_{\pi^0}$ are the cross sections for the
production of charged and neutral pions, respectively, in proton-proton collisions, and
the functions $F_e$ and $F_g$ are given by

\begin{equation}
F_e(\gamma_e,\gamma_p)=\int_{\gamma_\pi^{min}}^{\gamma_\pi^{max}} d\gamma_{\pi^\pm}
F_e(\gamma_e, \gamma_{\pi^\pm}) F_{\pi^\pm}(\gamma_{\pi^\pm}, \gamma_p)
\label{F_e}
\end{equation}
where $F_e(\gamma_e, \gamma_{\pi^\pm})$ is the electron energy distribution function
from pions, and $F_{\pi^\pm}(\gamma_{\pi^\pm}, \gamma_p)$ is the charged pion energy
distribution function;

\begin{equation}
F_g(E_g,\gamma_p)=\int_{E_\pi^{min}}^{\gamma_\pi^{max}} d\gamma_{\pi^0}
F_g(E_g, \gamma_{\pi^0}) F_{\pi^0}(\gamma_{\pi^0}, \gamma_p)
\label{F_g}
\end{equation}
where $F_g(E_g, \gamma_{\pi^0})$ is the photon energy distribution function
from pions and $F_{\pi^0}(\gamma_{\pi^0}, \gamma_p)$ is the neutral pion energy
distribution function. The distribution functions in (\ref{F_e}) and (\ref{F_g}) used here
are taken from Moskalenko \& Strong (1998).

Given a value of $\xi_0$, the $GR$ spectral brightness and
luminosity can be obtained immediately from integration of
(\ref{gprodrate}) out to a maximum radius $R$. For the $SRE$, in
order to obtain their number density as a function of energy and
radius, $n_e(\gamma_e, r)$, it is necessary to solve the equation
that locally governs (spatial diffusion can be regarded as
irrelevant in this context, see e. g. Sarazin 1999) the balance
between injection and energy losses,

\begin{equation}
\frac{\partial}{\partial \gamma_e} \left[ b(\gamma_e,r)
n_e(\gamma_e,r) \right] =-Q_e(\gamma_e,r)
\label{equilibrio}
\end{equation}
whose solution is given by

\begin{equation}
n_e(\gamma_e,r)=\frac{1}{b(\gamma_e,r)}\int_{\gamma_e} ^\infty
Q_e(\gamma_e',r)d\gamma_e'.
\label{soluzequilibrio}
\end{equation}

In (\ref {equilibrio}) and (\ref {soluzequilibrio}), $b(\gamma_e) \equiv -d\gamma_e /dt$
is proportional to the sum of the square of $B_{mwb}$, the magnetic field
corresponding to the energy density of the cosmic microwave background (IC losses)
and the square of $B(r)$. The latter quantity
is not fixed a priori, but will be derived through an iterative process,
when the $SRE$ is requested to reproduce the radio brightness
distribution for any given cluster. Then a radial behavior of
the magnetic field will univocally emerge, and the normalization
$B(r=0) \equiv B_0$ will follow from the adopted value of $\xi_0$.
As a reference value for the maximum in the
$GR$ emission to be expected, $\xi_0$ = 0.3 will be
adopted as representative of the most extreme conditions described
in Sect. 2 with regard to the $RP$ acceleration.

\section{Four clusters and the brightness profile of their radio halos}

The exercise is applied to four clusters as examples. Their
names and properties are summarized in Tables \ref{tab.par.4amm.1}
and \ref{tab.par.4amm.2}. In Table \ref{tab.par.4amm.1}, in a
sequence, after redshift and temperature, the three parameters
$n_{th,0}$, $r_{c,th}$, and $q_{th}$ are given, which
quantitatively define the
density distribution $n_{th}$, according to (\ref {radialthermal}). 
As in Paper I, the radial
(azimuthally averaged) brightness distribution at frequency $\nu$
was fitted with a functional form similar to the one usually
adopted for the X-ray brightness, namely (the subscripts $R$ here
mean ``radio'')

\begin{equation}
S_\nu(\theta)=S_{\nu,0} \bigg[1+ \bigg({\theta \over \theta_{c,R}}
\bigg)^2 \bigg]^{-q_R'},
 \label{bril.inf}
\end{equation}
where $\theta$ is the angular distance from the center. Under the
assumption of spherical symmetry, the profile of the radio
emissivity is similar in form to $n_{th}$ (\ref
{radialthermal}), with $q_R=q_R'+1/2$ and $r_{c,R}=
\theta_{c,R}D$, $D$ being the cluster angular distance. These are
the two quantities measured at $\nu$ = 1.4 GHz in Table
\ref{tab.par.4amm.2}, along with the flux at the same frequency,
the radius $R_h$, and the angular size $\theta_h$ of the
radio halo. The spectral index  $\alpha_R$ in the second
column is fairly well-determined only for the Coma cluster; for
the three other objects, the value is given with a double dot, to
remind the reader that it has not been measured well yet (the flux
density data available will appear in Figs. 1--4). Note
that the value of $s$ = 2.7, chosen in Sect. 2 as a reference
energy spectral slope of the $RP$, at equilibrium yields a slope of the $SRE$ 
that almost exactly matches the radio spectral index
of Coma (Fig.\ref{fig.radiosecd_coma}).

\begin{table*}[tbp]
 \caption{\footnotesize{
The four clusters, x-ray properties}}
\begin{center}
\begin{tabular}{ccccccccc}
 \hline \hline
    & $z$ &  $kT$ & $n_{th,0}$  & $r_{c,th}$& $q_{th}$ \\
    &     & (keV) & ($h_{70}^{1/2}$ cm$^{-3}$) & ($h_{70}^{-1}$ kpc) &  & \\
 \hline
 Coma & 0.023 [1] &  8.2 [2] & $3.4\times10^{-3}$ [2] & 300 [2] & 1.125 [2]\\
A2319 & 0.0557 [1] &  9.12 [3] & $7.83\times10^{-3}$ [4] & 94 [5] &  0.765 [5]\\
A2255 & 0.0806 [1] &  7.3 [6] & $1.72\times10^{-3}$ [7] & 410 [8] & 1.11 [8]\\
A2163 & 0.203 [1] &  14.6 [9] & $7.87\times10^{-3}$ [9] & 220 [9] & 0.93 [9]\\
 \hline
 \end{tabular}
\end{center}
\footnotesize{References: [1] Struble \& Rood (1999); [2] Briel, Henry \& B\"ohringer (1992);
[3] Arnaud \& Evrard (1999); [4] Trevese, Cirimele \& De Simone (2000);
[5] Feretti, Giovannini \& B\"ohringer (1997b); [6] David et al. (1993);
[7] Jones \& Forman (1984); [8] Feretti et al. (1997a);
[9] Elbaz, Arnaud \& B\"ohringer (1995).}
 \label{tab.par.4amm.1}
 \end{table*}

\begin{table*}[tbp]
 \caption{\footnotesize{
The four clusters, parameters of the radio halo}}
\begin{center}
\begin{tabular}{ccccccccc}
 \hline \hline
    & $\alpha_R$ &  $F(1.4\mbox{ GHz})$ & $r_{c,R}$& $q_{R}$ & $R_h$ & $\theta_h$ \\
    &     & (Jy) & ($h_{70}^{-1}$ kpc) &  & ($h_{70}^{-1}$ kpc) & (arcmin)\\
 \hline
 Coma & 1.35 [1] &  0.64 [1] & 670 & 4.4 & 900 [1] & 64 \\
A2319 & 1.8:  [2] &  0.153 [2] & 37 & 0.98 & 470 [2] & 16 \\
A2255 & 1.7:  [3] &  0.043 [3] & 470 & 3.0 & 670 [3] & 16 \\
A2163 & 1.18: [4] &  0.155 [5] & 220 & 1.4 & 1,050 [4] & 12 \\
 \hline
 \end{tabular}
\end{center}
\footnotesize{References: 
[1] Deiss et al. (1997); [2] Feretti et al. (1997b);
[3] Feretti et al. (1997a); [4] Feretti et al. (2004); [5] Feretti
et al. (2001).}
 \label{tab.par.4amm.2}
 \end{table*}

\section{Numerical results}

\begin{figure}[tbp]
\begin{center}
 \resizebox{\hsize}{!}{\includegraphics{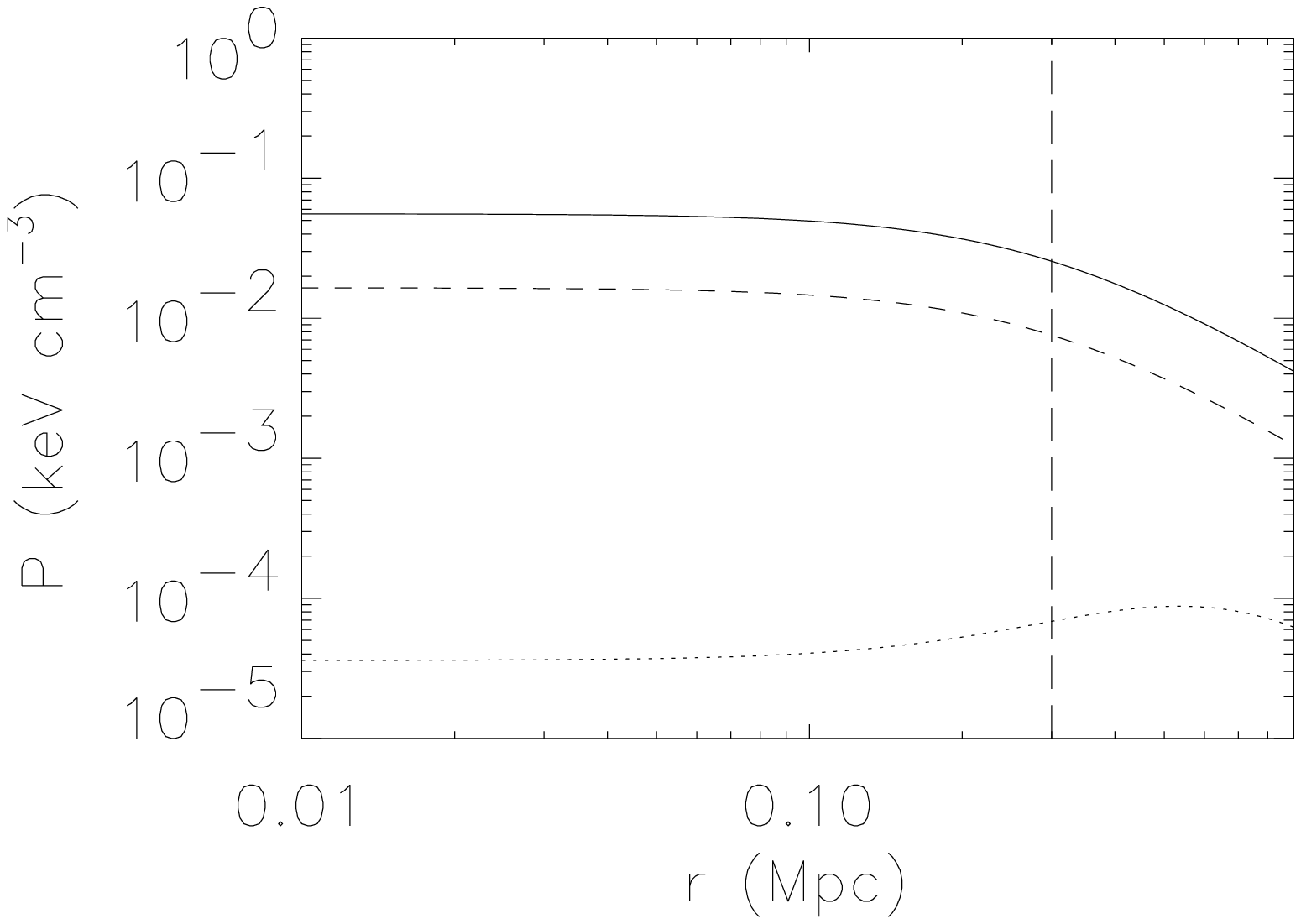}}
 \resizebox{\hsize}{!}{\includegraphics{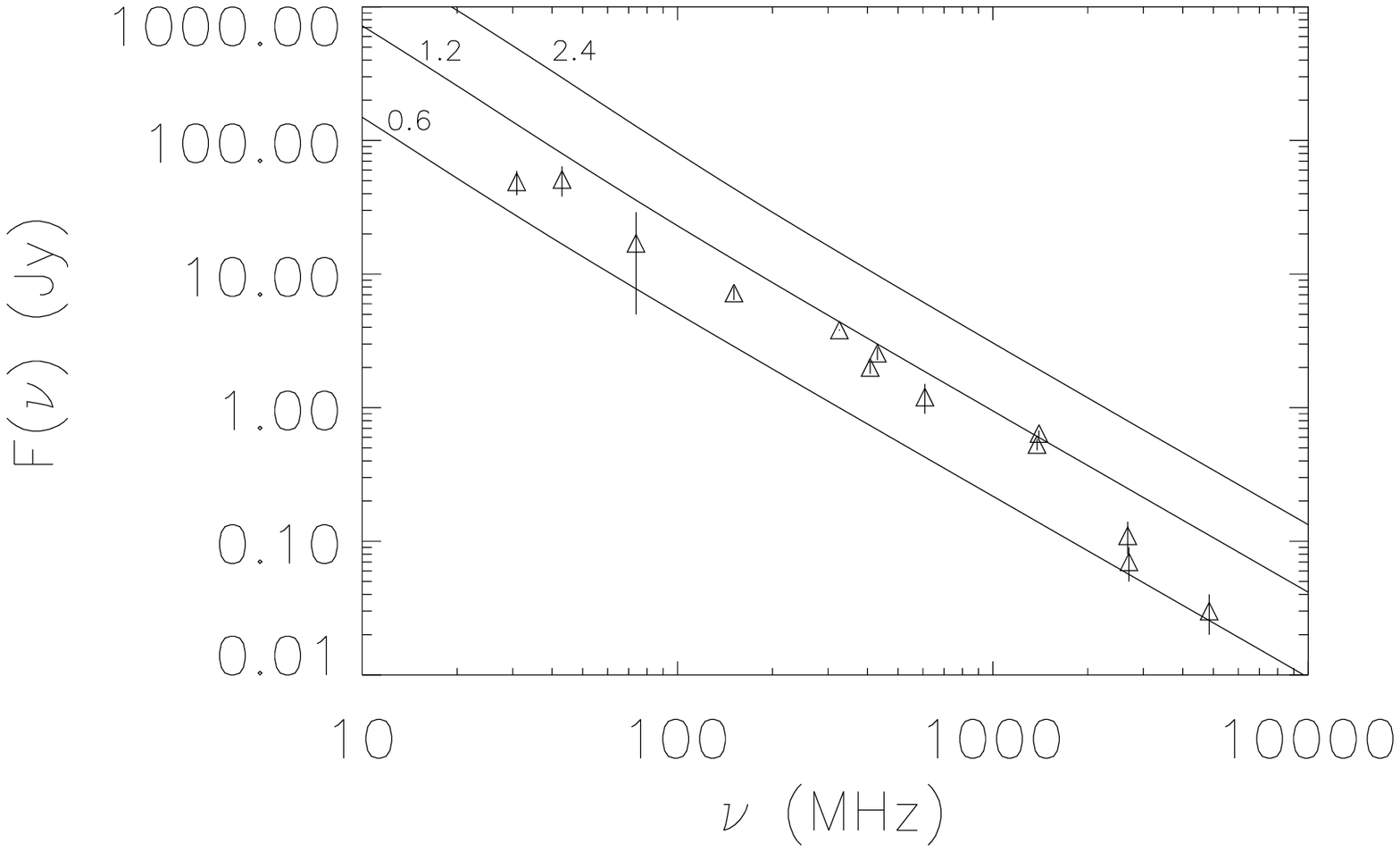}}
  \caption{\footnotesize{Coma cluster, $\xi_0 = 0.3$, $s=2.7$.
Upper panel: $P_{th}$ (full line),$P_{RP}$ (dashed line), and
the magnetic field pressure (dotted line), with $B_0=1.2$ $\mu$G,
as a function of distance from the center. The vertical dashed line
marks the position of $r_{c,th}$. Lower panel: the $SRE$ radio
emission spectra labelled with three values $B_0$ (in $\mu$G).
The data points are from Thierbach, Klein \&
Wielebinski (2003).
  }}\label{fig.radiosecd_coma}
\end{center}
\end{figure}

\begin{figure}[tbp]
\begin{center}
 \resizebox{\hsize}{!}{\includegraphics{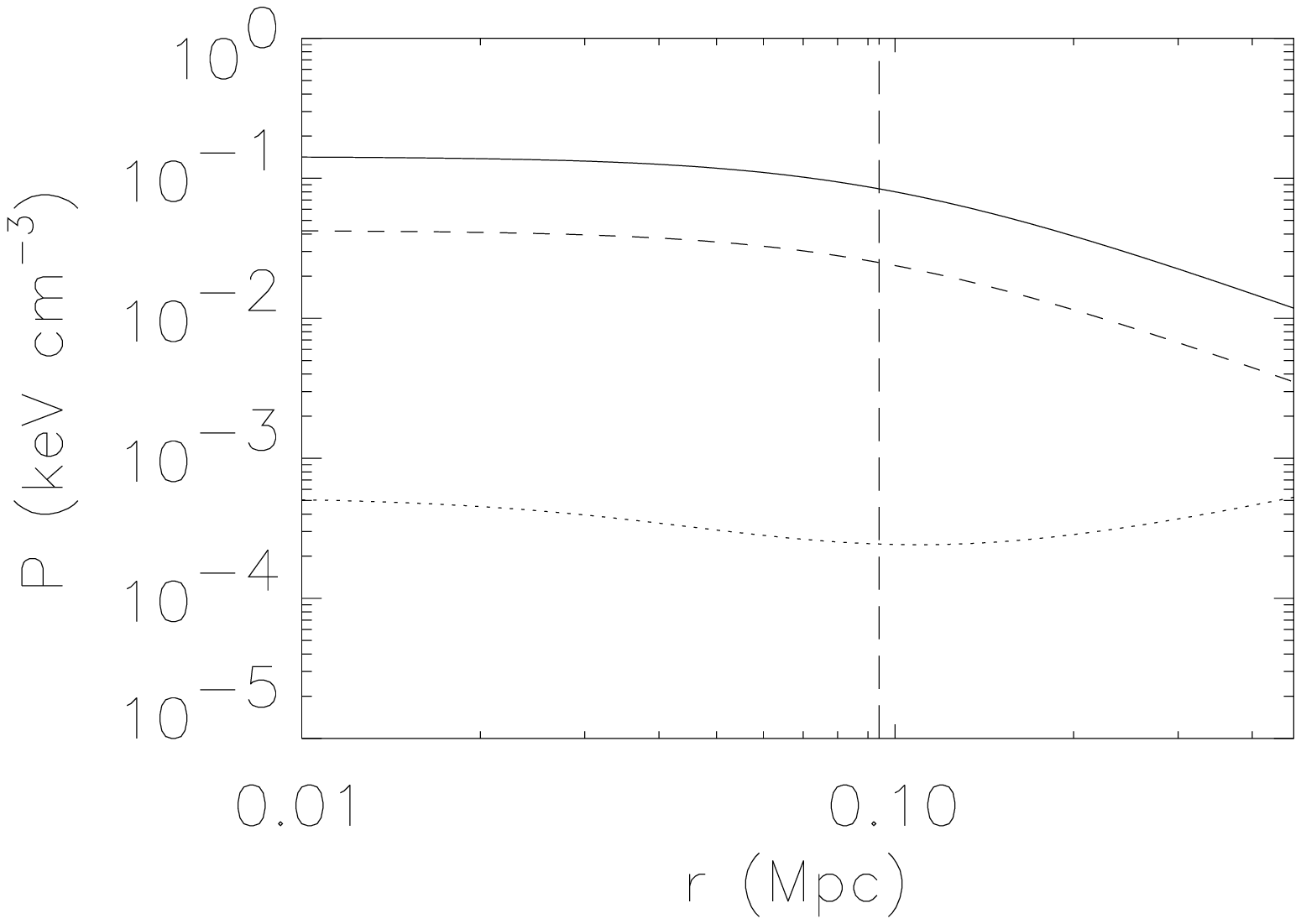}}
 \resizebox{\hsize}{!}{\includegraphics{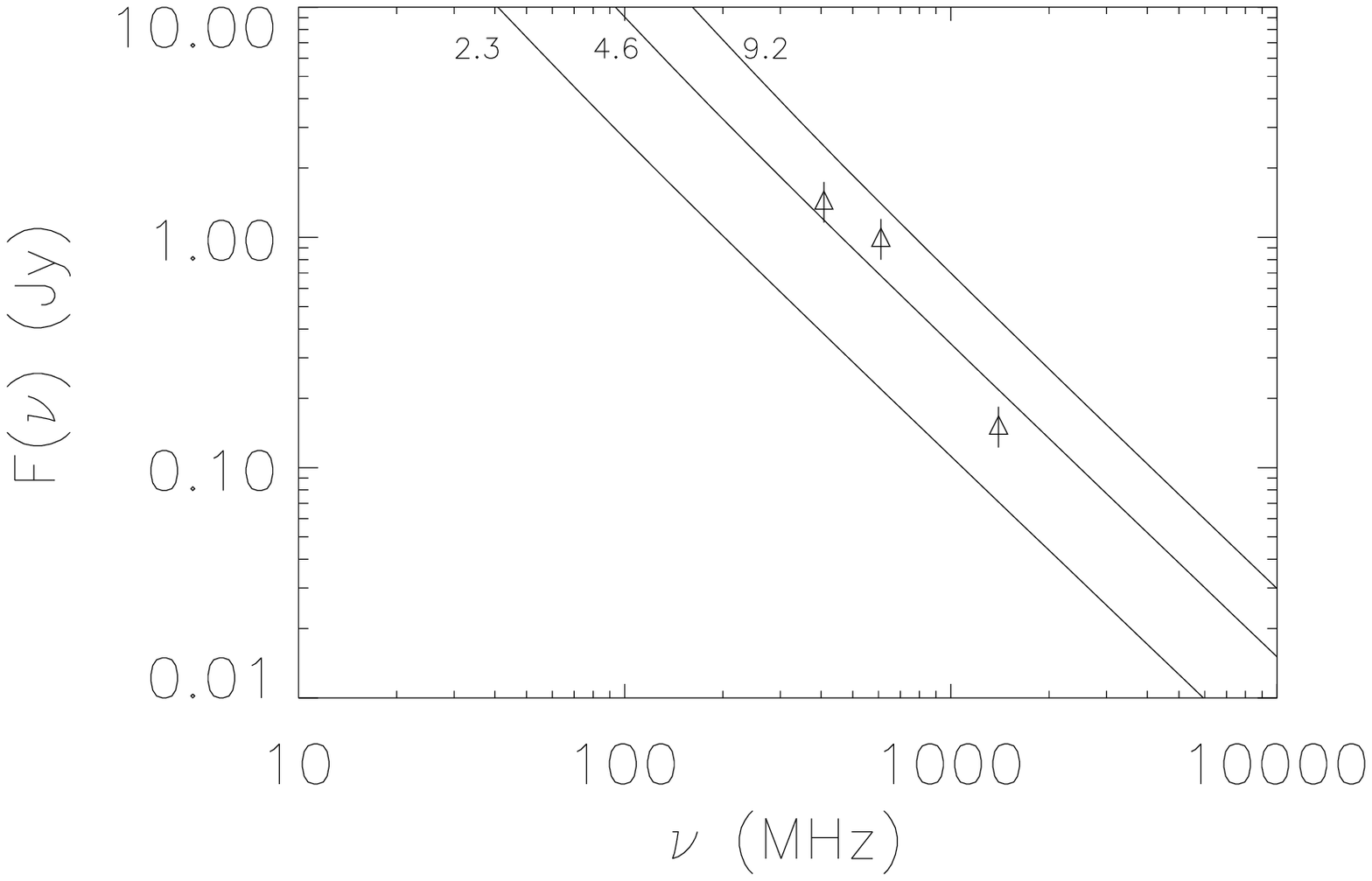}}
  \caption{\footnotesize{A2319. Same as Fig. 1, except for:
upper panel, $B_0=4.6$ $\mu$G. Lower panel, data without error bars from
Feretti et al. (1997b), with 20\% error bars added arbitrarily.
  }}\label{fig.radiosecd_a2319}
\end{center}
\end{figure}

\begin{figure}[tbp]
\begin{center}
 \resizebox{\hsize}{!}{\includegraphics{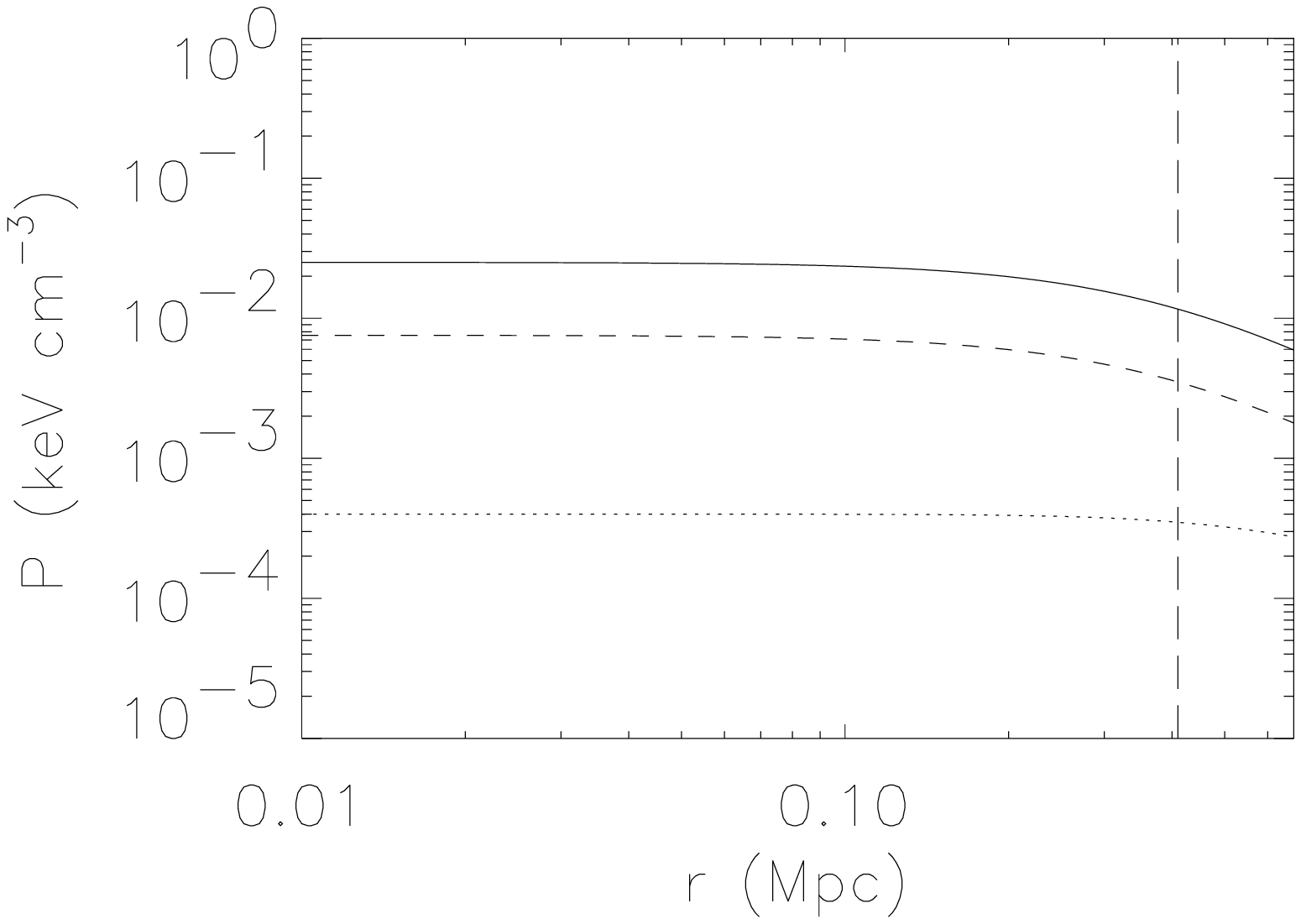}}
 \resizebox{\hsize}{!}{\includegraphics{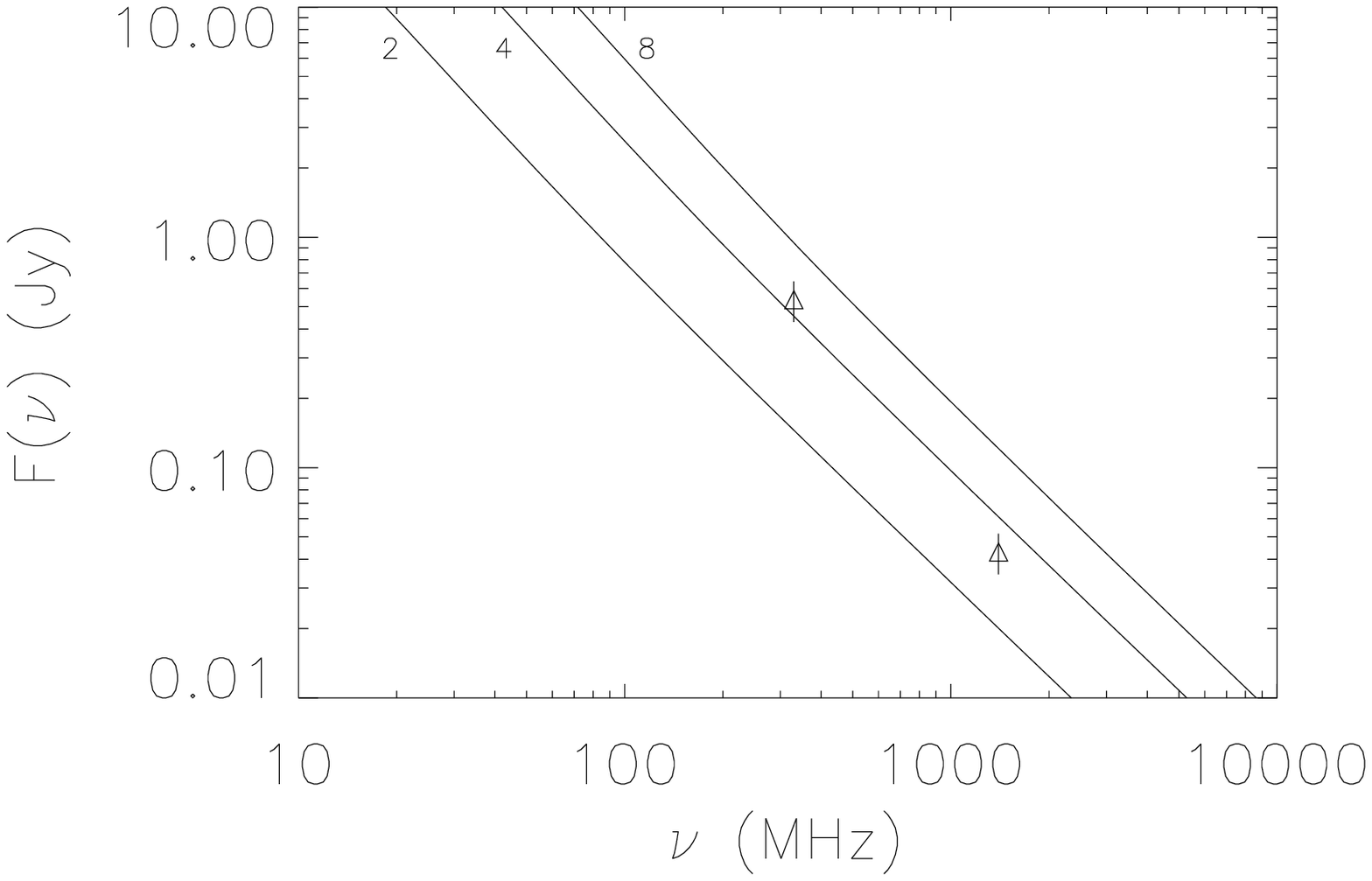}}
  \caption{\footnotesize{A2255. Same as Fig. 1, except for:
upper panel, $B_0=4$ $\mu$G. Lower panel, data without error bars from
Feretti et al. (1997a), with 20\% error bars added arbitrarily.
  }}\label{fig.radiosecd_a2255}
\end{center}
\end{figure}

\begin{figure}[tbp]
\begin{center}
 \resizebox{\hsize}{!}{\includegraphics{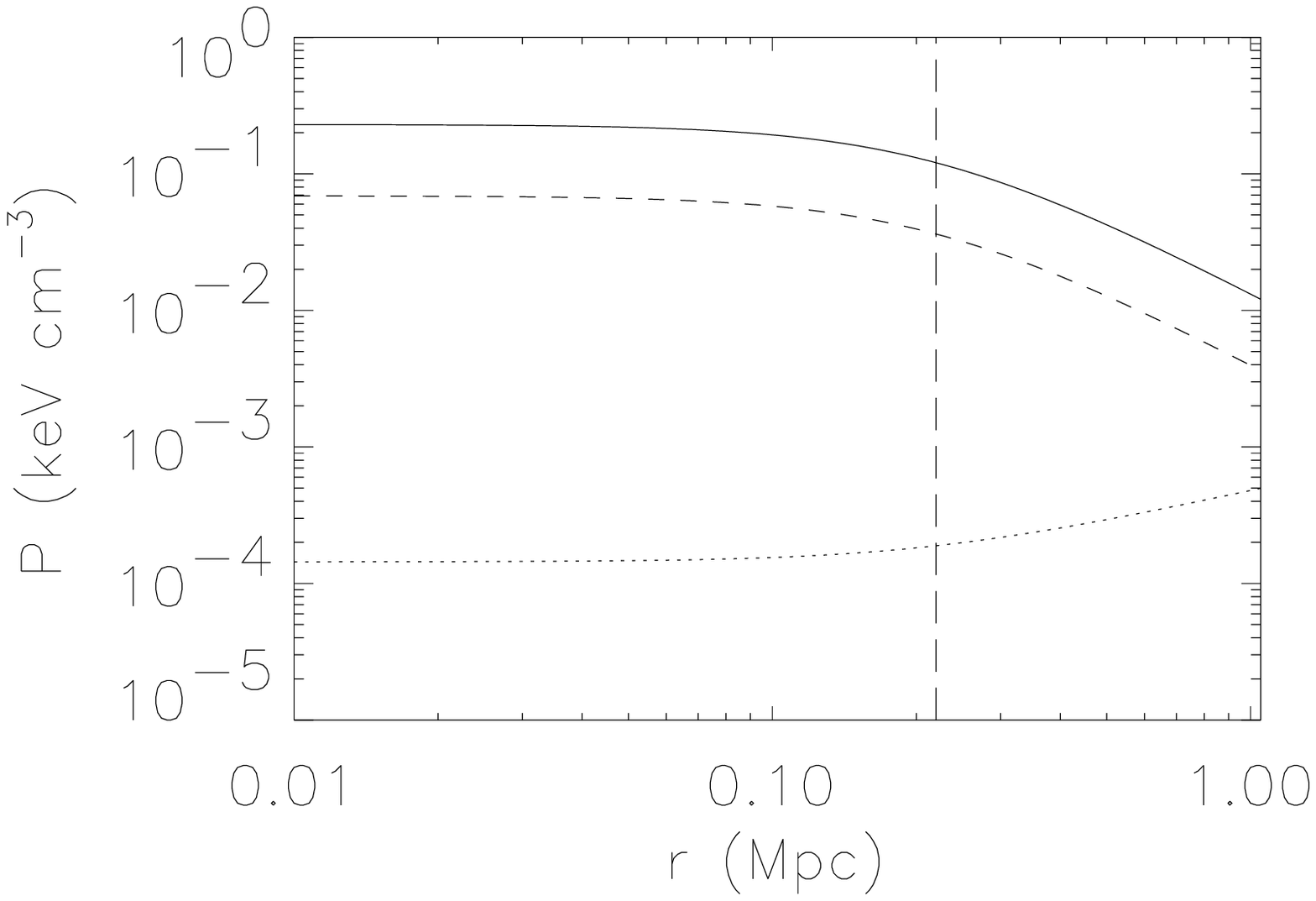}}
 \resizebox{\hsize}{!}{\includegraphics{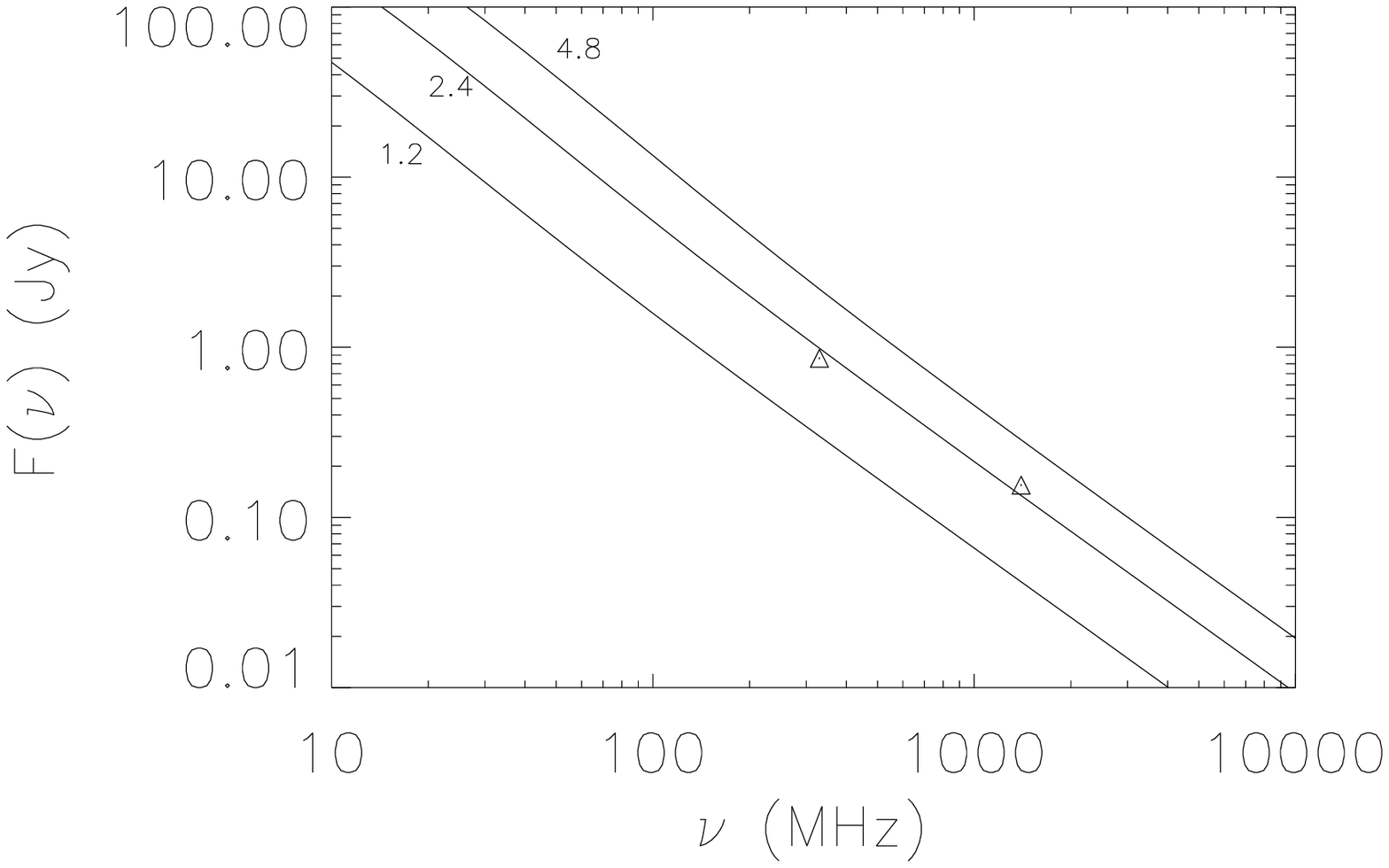}}
  \caption{\footnotesize{A2163. Same as Fig. 1, except for:
upper panel, $B_0=2.4$ $\mu$G. Lower panel, data from Feretti et al. (2001; 2004),
whose error bars are within the height of the two triangles.
  }}\label{fig.radiosecd_a2163}
\end{center}
\end{figure}

\begin{figure}[tbp]
\begin{center}
     \resizebox{\hsize}{!}{\includegraphics{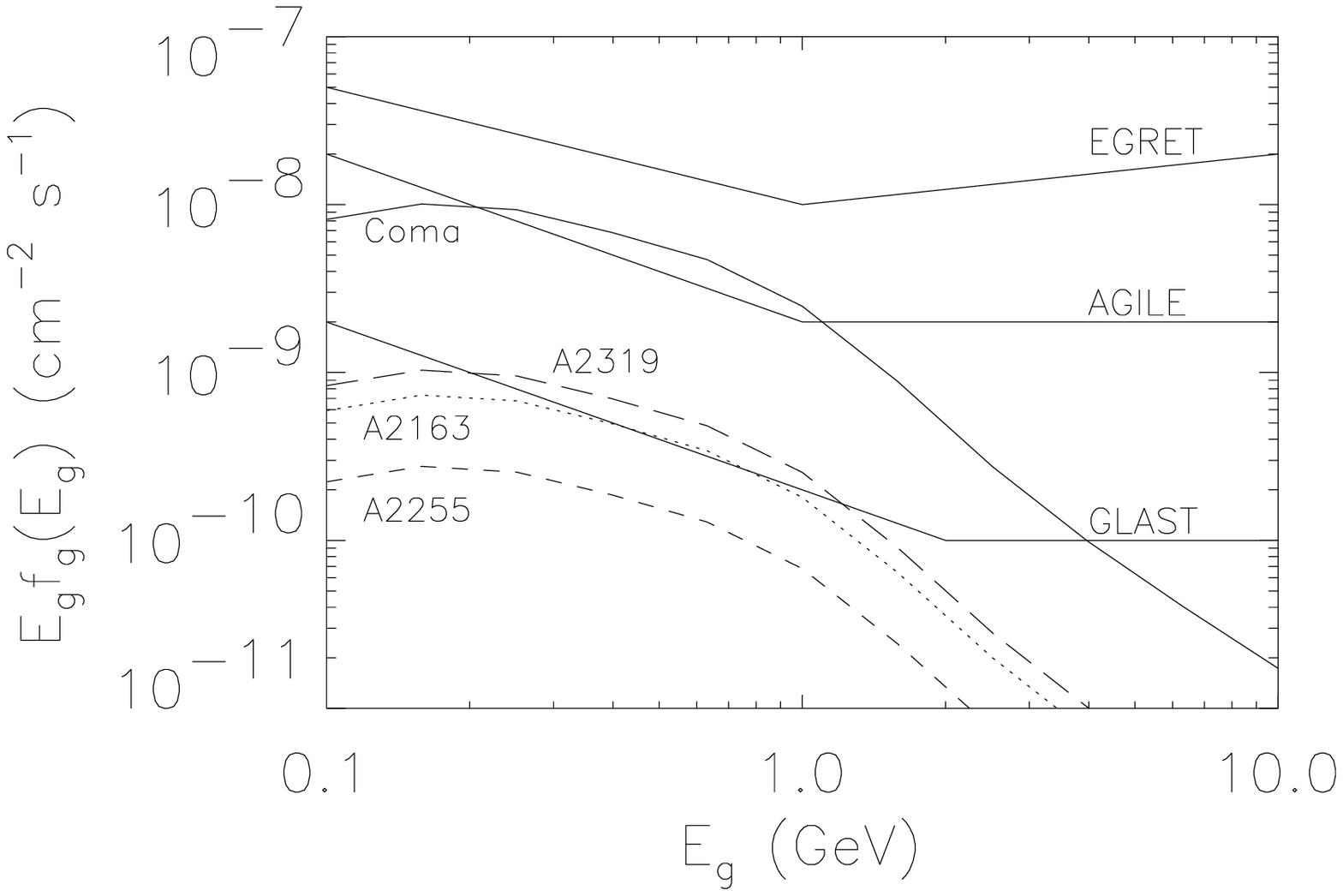}}
  \caption{\footnotesize{
Spectral energy distribution of the gamma ray flux
from $\pi^0$ decay expected in the case illustrated for
each cluster in Figs. 1 to 4. The sensitivity of
EGRET, AGILE, and GLAST is drawn for comparison purposes
and corresponds to a $5\sigma$ detection in a full-year sky survey
(from http://people.roma2.infn.it/$\sim$agile/).
   }}\label{fig.gammasec.4amm.d}
\end{center}
\end{figure}

The results are summarized in Table \ref{tab.par.risultati} and
illustrated in Figs. \ref{fig.radiosecd_coma},
\ref{fig.radiosecd_a2319}, \ref{fig.radiosecd_a2255},
\ref{fig.radiosecd_a2163}, and \ref{fig.gammasec.4amm.d}.
The following quantities are given in Table
\ref{tab.par.risultati}: the
number density of the $RP$ at the center, $n_{RP,0}$, the value of
$B_0$ for which the radio flux is best reproduced by the $SRE$
alone, the gamma ray flux from  $\pi^0$ decay, $f_g$(100 MeV),
obtained integrating out to the radius of the radio--halo $R_h$,
the gamma ray luminosity $L_g$ in the band 0.1 to 10 GeV, and next
to it, for comparison, the radio luminosity approximated as
1.4(GHz)$\times$L(1.4 GHz). The quantity $\tilde{B}_0$ in the
fourth column is the reduced mean value of $B$ if ``small-scale''
scalar fluctuations (see Paper I) are admitted to exist on the
order of $<(\delta B)^2>/B^2=1$ throughout a cluster.

In the upper panels of Figs. \ref{fig.radiosecd_coma} to \ref{fig.radiosecd_a2163},
the radial behaviors of $P_{th}$, $P_{RP}$, and of the magnetic
field pressure are shown out to $R_h$. Note that the magnetic pressure
remains lower than the other two everywhere. In the lower panels of the same
figures, along with the observational data, the radio spectral flux density
is drawn for three values of $B_0$, the one given in Table \ref{tab.par.risultati}
and two more, to illustrate the sensitivity of the normalization to
this quantity.

In Fig. \ref{fig.gammasec.4amm.d} the spectral density
distributions of the gamma rays from $\pi^0$ decay of all four
clusters are given. Note that the expected flux scales linearly
with $\xi_0$, and the values in the figure correspond to
$\xi_0=0.3$. For comparison the sensitivity curves are shown for
the experiment EGRET onboard the satellite Compton-GRO and for the
future AGILE and GLAST missions, for an exposure time as reported
in the caption. Although the curves apply to point-like
sources, note that the angular resolution (on-axis) of GLAST is
about 60 arcmin at 1 GeV (from
http://www-glast.slac.stanford.edu), hence it should also apply to
the extended sources considered here (see Table 2 for the angular
size of the radio halos). It is evident that out of the four
clusters considered, only the Coma cluster emission could be
significantly detected by GLAST, even if the value of $\xi_0$ were
lower than 0.3, down to about 0.1. With $\xi_0$ = 0.3, Coma
could be marginally detected by AGILE, while A2319 and A2163 
could be marginally detected by GLAST.

The discussion of these results is postponed to Sect. 7,
after considering a case of emission, in the ultraviolet,
which has been proposed in the literature as due to $SRE$.

\begin{table*}[tbp]
 \caption{\footnotesize{Numerical results
}}
\begin{center}
\begin{tabular}{ccccccccc}
 \hline \hline
    & $n_{RP,0}$  & $B_0$    & $\tilde{B}_0$ & $f_g$(100 MeV) & $L_g$ (0.1--10 GeV) & $\nu L_R$ (1.4 GHz) \\
    & (cm$^{-3}$) & ($\mu$G) & ($\mu$G)      & (cm$^{-2}$ s$^{-1}$ GeV$^{-1}$) &  (erg s$^{-1}$) & (erg s$^{-1}$) \\
 \hline
 Coma & $1.1\times10^{-5}$ & 1.2 & 0.8 & $8.4\times10^{-8}$ & $1.3\times10^{43}$ & $9.9\times10^{39}$ \\
A2319 & $2.4\times10^{-5}$ & 4.6 & 3.2 & $8.6\times10^{-9}$ & $8.6\times10^{42}$ & $1.5\times10^{40}$ \\
A2255 & $6.0\times10^{-6}$ & 4.0 & 2.8 & $2.2\times10^{-9}$ & $5.4\times10^{42}$ & $9.7\times10^{39}$ \\
A2163 & $2.0\times10^{-5}$ & 2.4 & 1.7 & $6.0\times10^{-9}$ & $1.0\times10^{44}$ & $2.6\times10^{41}$ \\
 \hline
 \end{tabular}
\end{center}
 \label{tab.par.risultati}
 \end{table*}

\section{$SRE$ and the EUV excess in the Coma cluster}

Within 25 arcmin of the center of the Coma cluster, 
the EUVE satellite detected
an emission in the far UV band 130--180 eV,
as reported in Lieu et al. (1996) and Bowyer et al. (1999, 2004). It
exceeds the extrapolation of the thermal emission measured in the
X-rays. The origin of this excess is controversial (Lieu et al. 1996;
En\ss lin \& Biermann 1998; Brunetti et al. 2001; Bowyer et al. 2004), here
only the hypothesis put forward by Bowyer et al. (2004) (see also  Kuo et al. 2005
for a generalization to the clusters as a class) will be considered, namely that, in the
first place, the emission mechanism is IC on the microwave background and,
in the second, that the electrons have a secondary origin. The reason
for inserting this issue into the present paper follows from the fact
that the UV measurement, albeit essentially monochromatic and without spectral
information, is spatially resolved; hence, under the IC and spherical
symmetry assumptions, both the number and the radial distribution of
the electrons, whose energy must be $\gamma_e \sim 300$,
can be derived in absolute terms. The amount of parent $RP$ can therefore
be derived, and their pressure compared with the thermal pressure to
verify if the secondary origin of the electrons remains tenable.

The profile of the UV surface brightness $S_{UV}(\theta)$ can be
described with the functional form of Eq. (\ref {bril.inf}), and
it turns out (as pointed out by Bowyer et al. 2004) to be very similar
to that of the thermal X-ray emission. A best fit with
$\theta_{c,UV} = \theta_{c,X}$ = 10.5 arcmin yields an exponent
$q_{UV}' = 1.92 \pm 0.20$ ($\chi_{red}^2$ = 0.64 with 11 dof),
consistent within 1$\sigma$ with  $q_X'$ = 1.75 (Briel, Henry \&
B\"ohringer 1992). The emissivity is given (see Sect. 4) by a
functional form like (\ref {radialthermal}), with $r_{c,UV} =
r_{c,th}$ and $q_{UV} = q_{UV}' +1/2 = 2.42 \pm 0.20$: in the
following $q_{UV}$ = 2.42 will be used. Since the emissivity is
proportional to the number density of the electrons $n_e$ with
$\gamma_e$ about 300, this quantity will have the same radial
dependence:

\begin{equation}
n_e(r)=n_e(0) \left[ 1+ \left( \frac{r}{r_{c,th}} \right)^2
\right]^{-2.42}. \label{elettroni.euv}
\end{equation}

The $RE$ with $\gamma_e$ about 300 would emit synchrotron radiation
at $\nu \sim 0.4(B/\mu$G) MHz, an inaccessible window.
As pointed out already by Bowyer \& Bergh\"ofer (1998),
the radial behavior (\ref {elettroni.euv}) is likely to be very dissimilar from that of the
$RE$ responsible for the observed radio emission; otherwise, it would imply
a magnetic field increasing with $r$ out to at least 2/3 $R_h$ 
for the radio brightness distribution of Coma C.

Given the $SRE$ density distribution (\ref {elettroni.euv}), one
can resort to the population of the parent $RP$ by inverting the
procedure described in Sect. 4. The first step consists in
obtaining $Q_e(\gamma_e, r)$ from Eq. (\ref {equilibrio}).
Concerning the energy-loss rate, $b(\gamma_e, r)$, it now consists
of two terms, the Coulomb loss rate $b_c$, which depends on $r$
since $b_c \propto n_{th}$, and the IC loss rate $b_{IC}$, which
is constant in space and corresponds to the loss rate in an
equivalent magnetic field $B_{mwb} = 3(1+z)^2 \mu$G. The
synchrotron losses depend on the unknown value and radial
behavior of the magnetic field: ignoring these losses does not
affect the conclusion, as will be clear later on. Using the 
expression for $b_c$ given in Sarazin (1999), it turns out that
(for $n_{th}$ constant with time) at $r$ much less than $r_c$ and
for electrons with $\gamma_e \sim 300$,  $b_c$ was equal to
$b_{IC}$ at $z^*$ = 0.384, and their lifetime was 2.1 Gyr at that
epoch. The time elapsed since then up to the epoch corresponding
to the redshift of the Coma Cluster -- for $H_0=70$ km s$^{-1}$
Mpc$^{-1}$, $\Omega_m=0.3$, and $\Omega_\Lambda=0.7$ -- is 3.4 Gyr,
therefore the UV emission observed now from the innermost region
of the cluster is due to electrons affected more by Coulomb
than by IC losses. At larger distances from the center the latter
dominate instead.

Equation (\ref {equilibrio}) contains the spectral energy distribution of the
electrons, which is unknown. The calculations were carried out in full,
that is by using Eq.(\ref {equilibrio}), in combination with Eq.(\ref {elprodrate}),
adopting two different values of $s$ for the $RP$, namely 2.7 and 2.3.

\begin{figure}[tbp]
\begin{center}
    \resizebox{\hsize}{!}{\includegraphics{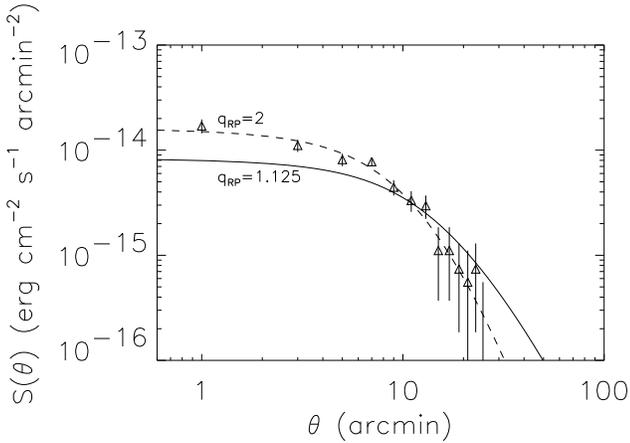}}
  \caption{\footnotesize{The UV brightness profile from $SRE$ for
two values of $q_{RP}$ (see text). The data are from Bowyer et al. (2004).
  }}\label{fig.fit_brill_euv}
\end{center}
\end{figure}

The radial distribution of the parent $RP$, which reproduces the UV brightness
profile fairly well (see Fig. \ref{fig.fit_brill_euv}), is given by

\begin{equation}
n_{RP}(r) \propto \left[ 1+ \left( \frac{r}{r_c} \right)^2 \right]^{-q_{RP}},
\label{protoni.euv}
\end{equation}
with $r_c = r_{c,th}$ and $q_{RP}$ = 2 (irrespective of which one of the two
values of $s$ is used). The exponent  $q_{RP}$ is
rather well constrained: as shown in Fig. \ref{fig.fit_brill_euv}, the UV brightness distribution
could not be reproduced if the $RP$
shared the same radial distribution with the thermal gas, namely if $q_{RP} = q_{th}$ = 1.125.
This would, notably, turn out to be the case if the Coulomb losses had been ignored
and only the spatially ``constant'' IC losses taken into account.
The result obtained implies that the quantity $\xi (r)$,
defined by (\ref {xi}) is not constant and that it is maximum at the center.

The normalization of Eq. (\ref {protoni.euv}) and the corresponding $P_{RP}$ depends on the exponent $s$.
The values of $\xi (r=0)$ for the two choices of $s$, obtained by integrating
the $RP$ spectrum down to the $\gamma_{min}$ given by Eq. (\ref {gammamin}), are
given in Table \ref{tab.euv.sec} and are much higher than unity
for both values of $q_{RP}$ in Fig. \ref {fig.fit_brill_euv}.
The inevitable conclusion, which would be even stronger
if the synchrotron losses were non-negligible, is that the hypothesis
of a secondary origin of the electrons is not tenable in this case.

 \begin{table}[tbp]
 \caption{\footnotesize{The value of $\xi (r=0)$, for
two choices of the exponents $s$ and $q_{RP}$, needed to
explain the UV excess in the Coma Cluster as due to
$SRE$
}}
\begin{center}
\begin{tabular}{ccc}
\hline \hline
$s=2.7$ &  $\xi (r=0)$ \\
 \hline
$q_{RP}=1.125$ & 9 \\
$q_{RP}=2$ & 21 \\
\hline
 \end{tabular}
\begin{tabular}{ccc}
\hline \hline
$s=2.3$ &  $\xi (r=0)$ \\
 \hline
$q_{RP}=1.125$ & 7 \\
$q_{RP}=2$ & 15\\
\hline
 \end{tabular}
 \end{center}
\label{tab.euv.sec}
 \end{table}

Surprisingly, Bowyer et al. (2004) reach the opposite conclusion. This
follows, partly, from their decision to ignore the Coulomb losses, 
in addition to the synchrotron losses. But this implies at most a difference 
in the value of $\xi$ by a factor five.
Since other approximations only have marginal effects, one
is led to conclude that their result is affected by a material error.
In this respect, it is worth mentioning that
Miniati et al. (2001b)
conclude that the UV excess due to $SRE$
should be lower than the measured value,
on the basis of a ratio 1/2 between the energy
content in $CR$ and in thermal gas obtained with numerical simulations
of the acceleration processes. 
All the more so, when account is taken
of the fact that the value adopted for comparison is the one reported by Bowyer et al. (1999),
which is a factor 2.7 lower than the one later presented in Bowyer et al. (2004).

\section{Discussion}

It is convenient to start by focusing the discussion on the central value
$B_0$ and the radial behavior of the magnetic field. The values of $B_0$
in Table \ref{tab.par.risultati} span an interval from 1.2 (Coma) to 4.6 $\mu$G (A2319). In all
four cases, the corresponding pressure is well below the $CR$ pressure.
If the ``true'' magnetic field strength were {\it substantially} smaller
or larger, the consequence for the $SRE$ in the central region would be,
in the first case, that their contribution must be negligible (otherwise
$\xi_0$ should be unacceptably large), in the second case that their contribution
is even more likely to be important, since $\xi_0$ could take a value
lower than 0.3, which is the reference value adopted above to approximately represente
the ``maximum'' $RP$ content. A comparison with the current estimates of $B_0$
from FR measurements is therefore in order.

None of the values of $B_0$ in Table \ref{tab.par.risultati} seems to be
inconsistent with typical FR estimates in the inner regions
of clusters, but one must take into account (see Paper I and references
therein for a brief presentation of this issue)
the rather large uncertainties affecting the FR results, which are
mainly related to the approximations adopted for the spectrum of the field
fluctuations as a function of their scale length. In particular, the
value for the Coma cluster is in fair agreement with the estimate obtained by
Kim et al. (1990) using background sources, but not with the one obtained by Feretti et al. (1995),
using one extended source within the cluster, which could be as large as $\sim$ 8 $\mu$G, if a
very small scale length for field reversals
is adopted, of the order of 1 kpc. The value
recently derived for A2255 (Govoni et al. 2006), using three extended sources
within the cluster and a more sophisticated approach than in Feretti et al. (1995),
is 2.5 $\mu$G (the confidence level is not explicitely
given in the paper), which is not far from the 4.0 $\mu$G, and
practically equal to $\tilde{B}_0$, in Table \ref{tab.par.risultati}.
For the two other clusters the ``individual''
FR measurements available are not reliable.
It must, in addition, be noticed that, in the frequency range around 1.4 GHz,
the $SRE$ spectral energy slope is
about 3.7 (for the choice made of $s$ = 2.7), because of the radiative losses,
and the radio emissivity
scales as $B^{(3.7+1)/2}$ = $B^{2.35}$: thus, it is very sensitive to
the $B$ value, as also illustrated in Figs. \ref{fig.radiosecd_coma}
to \ref{fig.radiosecd_a2163}.
Therefore, so long as a very robust FR estimate for each individual cluster
is lacking, a generic agreement is by no means a strong enough argument for
concluding that the $SRE$ might always be an important, if not
the dominant, contribution. Moreover, a very relevant uncertainty
on the values of $B$ for the three clusters other than Coma stems
from the large uncertainty on the radio spectral slope, hence on
their specific value of $s$ and consequently on the $RP$ pressure
for a fixed value of $\xi_0$ (Eqs. \ref {eq.spettro.protoni},
\ref {eq.press.cr}, and \ref {xi}). At present, it can only
be stated that the above-mentioned conclusion cannot be excluded
for the inner region of a cluster and its radio halo.

\begin{figure}[tbp]
\begin{center}
 \resizebox{\hsize}{!}{\includegraphics{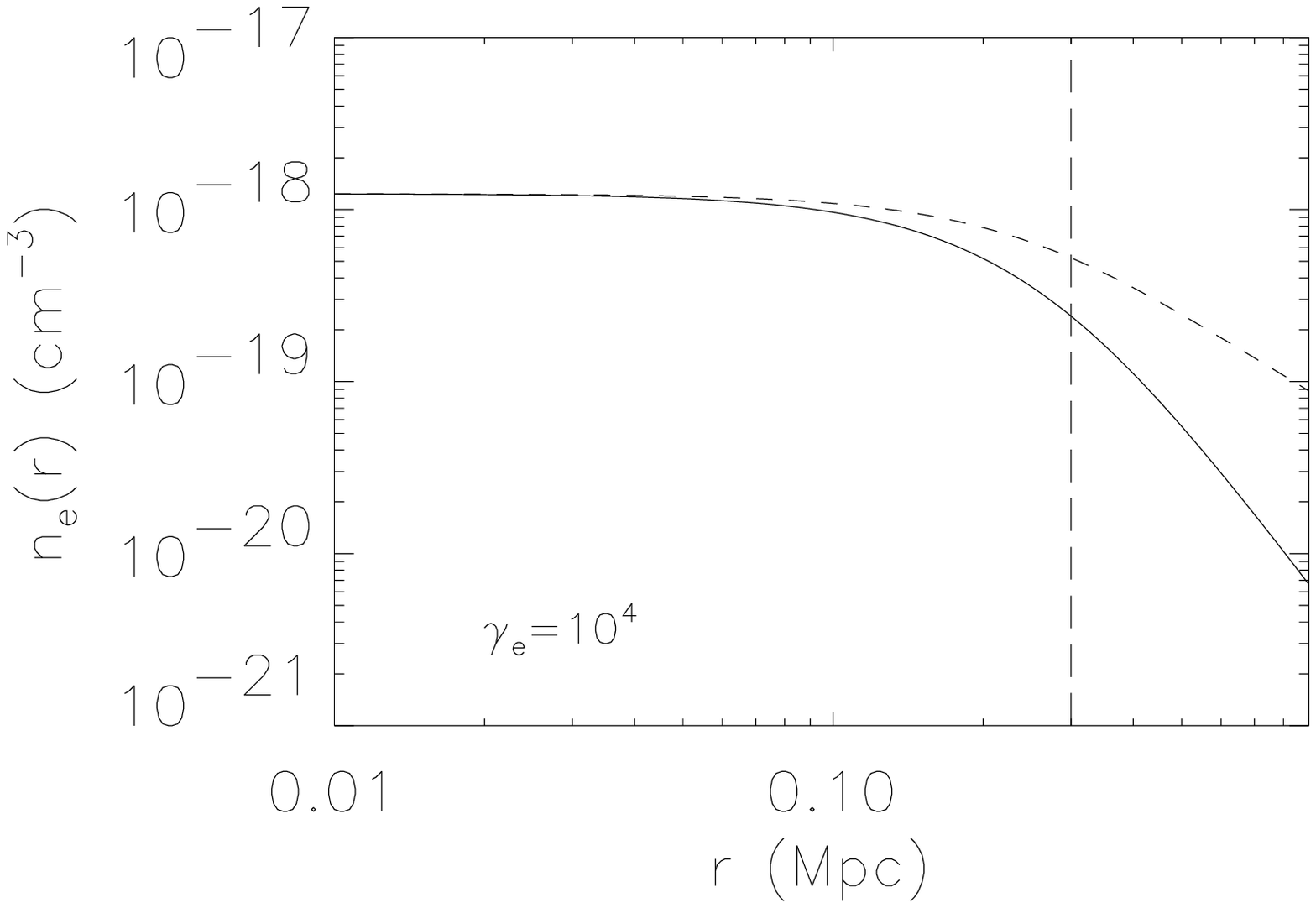}}
 \resizebox{\hsize}{!}{\includegraphics{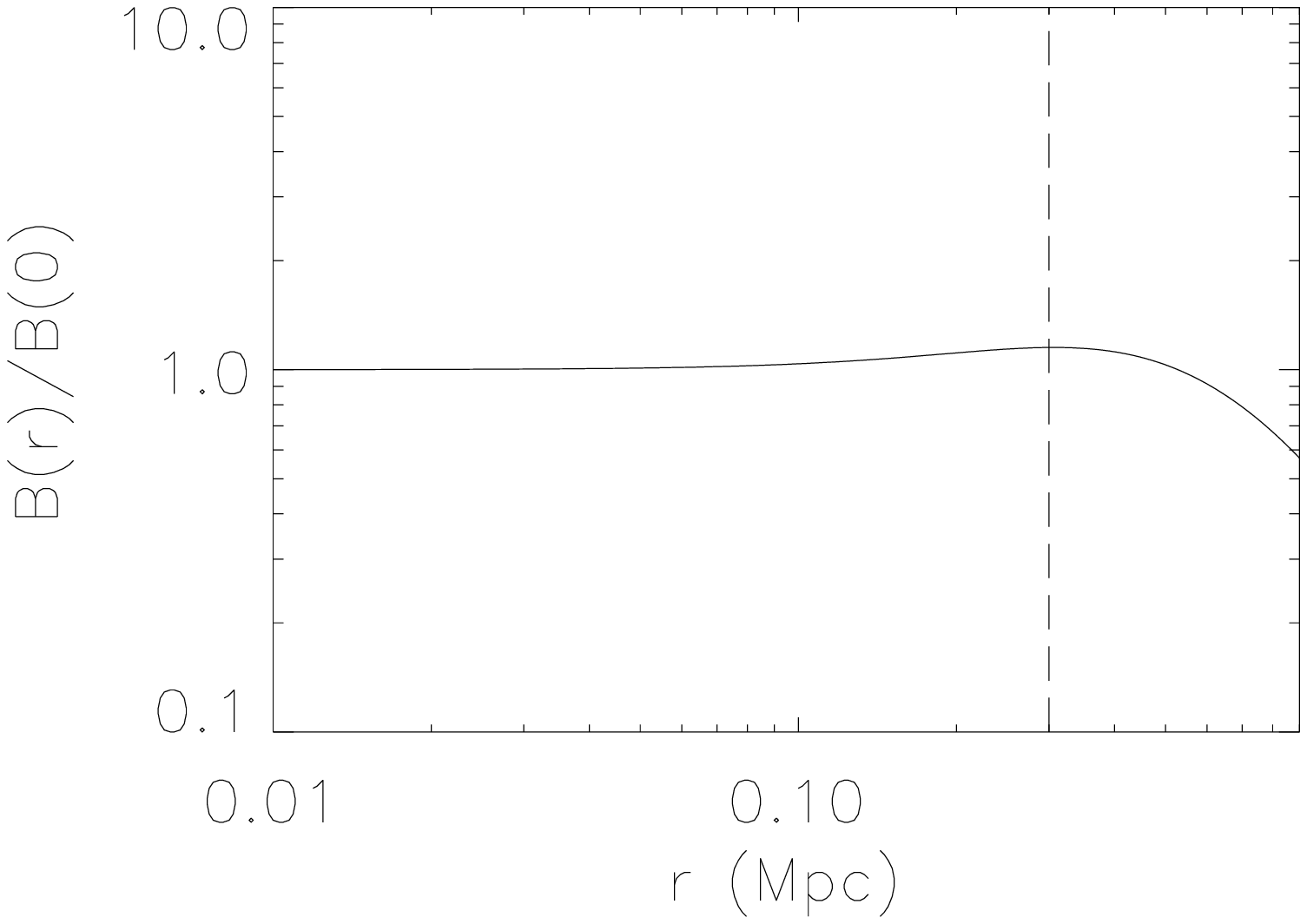}}
  \caption{\footnotesize{Upper panel: hypothetical mix
of $SRE$, in the case $\xi_0=0.15$, with $PRE$ of equal density at
the center of the Coma Cluster. The density refers to
electrons with $\gamma_e = 10^4$, the energy corresponding
approximately to the synchrotron frequency $\nu_c$=1.4 GHz when
$B_0$=1.2 $\mu$G. Lower panel: the radial behavior of the
magnetic field required to reproduce the radio surface brightness.
See text for details.
  }}\label{fig.misto.coma}
\end{center}
\end{figure}

The situation is plainly different at distances from the center
that are much greater than $r_{c,th}$. In the three objects where $R_h$ is
several times larger than the core radius, Figs.
\ref{fig.radiosecd_coma}, \ref{fig.radiosecd_a2319}, and
\ref{fig.radiosecd_a2163} show that the magnetic pressure,
although it remains well below the $CR$ pressure, tends to {\it
increase}, a behavior that is physically embarrassing.
Rather than sharply concluding that the $SRE$ should therefore
be regarded as a minor component of the bulk $RE$, it is perhaps more
sensible to consider the possibility that the contribution
of the $PRE$ might become dominant at large radii, thus
providing a way to get around this problem. To qualify this
possibility, one can, for instance, assume a continuous injection
of fresh $PRE$ throughout the cluster, described by a source function
$Q_{ep}(\gamma_e,r)$,
with a radial dependence of the form (8) adopted above for the
accumulated $RP$ and an energy dependence with the same index $s$.
The solution (14) would then give the equilibrium distribution,
whose energy dependence is practically the same as that of the
$SRE$. One of the results for the case of Coma is illustrated
in  Fig. \ref{fig.misto.coma}. Here the $PRE$ and $SRE$ contributions
at the center are assumed to have equal weight (that is, the
quantity $\xi_0$ was reduced to 0.15 to ensure that the $SRE$
central density amounts to half the value derived previously, while
the value of $B_0$ obviously remains the same, 1.2 $\mu$G).
The rise in magnetic pressure
present in Fig. \ref{fig.radiosecd_coma} has gone away almost
completely, except for the mild bump around 400 kpc, which is
evidently the artificial consequence of $r_{c,R}$ being about
twice $r_{c,th}$ (see Tables \ref {tab.par.4amm.1} and
\ref{tab.par.4amm.2}), together with the prescription $r_{c,RP} =
r_{c,th}$ adopted in Sect. 3.
In conclusion, to obtain an acceptable radial behavior of the
magnetic field, one must resort to $PRE$ at large distances from
the center, but they need not necessarily be the dominant
fraction within $r_{c,th}$, where the production of $SRE$ is most effective.

Observationally, the case of the UV excess analyzed in Sect. 6
is very instructive. If the surface brightness of the IC emission by the electrons
responsible for the radio emission could be measured in the hard X-ray band, the same
procedure could be followed to verify whether or not the amount
of parent $RP$ is compatible with the condition $\xi$ less than unity 
everywhere throughout the cluster. A positive answer would not,
however, be conclusive as to the secondary origin of the $RE$.
This issue
can only be addressed directly through measurements of the
collisionally-produced $GR$. This brings one back to the
key issue of the effective $CR$ content of clusters.

In this respect, a final important remark is due on the apparent
dichotomy between those clusters endowed with a radio halo and
those that are not. In the current literature (Feretti, Burigana
\& En\ss lin 2004), a correlation is emphasized between the
existence of a radio halo and the evidence of ongoing merging
phenomena. This correlation speaks in favor of a dominant
population of $PRE$ everywhere in these clusters, which are being
accelerated since at most one billion years. However, it must be
stressed that an attempt to estimate the strength of the magnetic
field in those clusters {\it devoid} of a diffuse radio emission by
combining FR measurements in an ensemble of them (Clarke, Kronberg
\& B\"ohringer 2001) 
resulted in a value of $B$ in the interval 1
to 10 $\mu$G, the same interval of the clusters with a radio halo. 
The results 
obtained in Sect. 5 would then suggest that 
the amount of RP, which have accumulated
in \textit{both} types of clusters
all along their ten billion years history,
should be significantly lower than the amount corresponding
to the reference value $\xi_0$ = 0.3.
It is the opinion of the authors of
the present paper that more sensitive measurements of a diffuse
radio emission in clusters, where it has gone undetected so far
(along with more accurate ``individual'' estimates of $B$ using
the FR), might provide very stringent constraints on their $CR$
content. 
In consideration of the hard fact that
the detection of $GR$ from more
than just one cluster will be a difficult task even for the mission GLAST,
further efforts should be made in the radio window.
In order to stimulate such efforts,
some clusters, where a diffuse radio emission has gone undetected so far,
and which look like promising targets, 
have been selected (Table \ref{tab.radioflussi.altri}, lower part). 
For reasons of homogeneity with the ones
discussed in this paper, none of these clusters shows the soft
excess usually named ``cooling flow''. 
The diffuse radio emission
from these clusters, 
which should be due exclusively to $SRE$, 
should be typically more concentrated than in the
``radio halo'' clusters,
because of the dependence of the production rate 
from the product $n_{th}\times n_{RP}$. 
Thus, the 1.4 GHz flux (Table \ref{tab.radioflussi.altri}, col. 3) was calculated
(according to the prescriptions in Sect. 3) for each of them and on the basis
of its X-ray properties, from within
$r=r_{c,th}$ (Table \ref{tab.radioflussi.altri}, col. 2), assuming the reference value 0.3 for $\xi_0$, and
a value equal to 5 $\mu$G for $B_0$. Note that the predicted flux values
scale linearly with $\xi_0$ and, in first approximation,
with the square of $B_0$. In the same table, for comparison, the
flux is reported from the corresponding region of the four
clusters studied above. Despite the large uncertainties associated
with the actual value of $B_0$, it is encouraging that the
predictions are comparable, in order of magnitude, with the
measured values of the four clusters. Furthermore, as reported in
the same table (col. 4), interesting upper limits already exist for three of them.
One case seems particularly significant, namely
A119, where the prediction and the upper limits are comparable,
while some FR measurements are available (Clarke, Kronberg \&
B\"ohringer 2001; Feretti et al. 1999) that indicate a $B_0$ value
of about 5 $\mu$G in this cluster (Murgia et al. 2004). Thus an
improvement by a factor of a few in the radio upper limit could
already place a significant constraint on $\xi_0$. Also the case
of A399, where the upper limit is about ten times lower than the
prediction, could be very significant, but the authors are not
aware of reliable FR measurements for this object. The last
column of Table \ref {tab.radioflussi.altri} reports the predicted
gamma-ray fluxes in the $0.1-10$ GeV interval from within
$r=r_{c,th}$: from a comparison with the similarly defined fluxes
of the four clusters in the upper part of the table, they appear
to be hardly detectable, with the possible exception of A3667.

  \begin{table*}[tbp]
 \caption{\footnotesize{The four clusters studied in this paper
(upper part) and a sample of six clusters whitout a radio halo
(lower part)}}
\begin{center}
\begin{tabular}{ccccc}
\hline \hline
Cluster &   $r_{c,th}$ & $F_{1.4}(r\leq r_{c,th})$ & $F_{1.4}(r\leq r_{c,th})$ & $F_g(0.1-10 \mbox{ GeV}, r\leq r_{c,th})$ \\
        &  (arcmin) & (mJy) &  (upper limit, mJy) & (cm$^{-2}$ s$^{-1}$)\\
 \hline
Coma & 10.5 & 160 &  & $6.8\times10^{-9}$\\
A2319 & 1.6 & 18  & & $2.7\times10^{-10}$\\
A2255 & 4.8 & 25  & & $2.7\times10^{-10}$\\
A2163 & 1.2 & 18 & & $2.9\times10^{-10}$\\
\hline
A119 & 5.3 & 30 & $\leq$ 47 & $1.8\times10^{-10}$\\
A2063 & 1.3 & 6 & & $3.8\times10^{-11}$\\
A1775 & 1.8 & 6 & & $4.1\times10^{-11}$\\
A1413 & 0.7 & 8 & & $5.8\times10^{-11}$\\
A399 & 1.9 & 22 & $\leq$ 1.6 & $1.5\times10^{-10}$\\
A3667 & 7.9 & 54 & $\leq$ 209 & $3.3\times10^{-10}$\\
\hline
 \end{tabular}
 \end{center}
\footnotesize{The X-ray parameters adopted for the predictions in the lower part come
from: Cirimele, Nesci \& Trevese (1997) for A119, A2063, A1775 and A1413;
 Sakelliou \& Ponman (2004) for A399; Vikhlinin, Markevitch
 \& Murray (2001) for A3667. The flux upper limits in A119, A399
derive from Giovannini et al. (2006), the one in A3667 from Feretti (2004).}
\label{tab.radioflussi.altri}
 \end{table*}

\section{Conclusions}

The issue of the CR content in galaxy clusters, which
are endowed with a diffuse radio emission called radio halo, was
discussed here in terms of the secondary products of collisions
with the thermal gas, namely $SRE$ and
$GR$. This production is maximized when the density
profile of the $RP$ is identical in form to
that of the thermal gas. The production rate can be normalized by
adopting a value of the ratio between $RP$ and thermal pressure,
$\xi_0$, constant throughout the cluster out to the radius $R_h$
of the radio halo. This quantity represents the efficiency and
output of current models of $CR$ acceleration associated to
merging processes in clusters. The value adopted here as a
reference is $\xi_0=0.3$ to represent the case of maximum
efficiency and cumulation over the cluster lifetime. A
representative value $s$ = 2.7 has been chosen for the energy
spectral index of the $RP$.

The equilibrium spectrum of the $SRE$ is then constrained to
reproduce the observed radio emission, and both the central value
$B_0$ and the radial behavior of the magnetic field is then
uniquely derived. In the four clusters treated, the values of $B_0$
fall between 1.2 and 4.6 $\mu$G (reduced to 0.8 to 3.2 $\mu$G if
scalar fluctuations on the order of $<(\delta B)^2>/B^2=1$ are
admitted): these values are consistent with estimates
of the same quantity from FR measurements
within clusters in general,
but more detailed FR measurements are needed for each cluster
individually before a sound conclusion might be drawn.
Particularly in view of the high sensitivity of the synchrotron
emissivity to the value of $B$.

The radial behavior of $B$,
instead, seems unnatural, in that it tends to increase with the radial distance from
the center. Rather than definitely exclude a relevant contribution of the $SRE$
to the radio emission also in the central parts of a cluster, this result suggests
that $PRE$, accelerated ``recently'', i.e. within their maximum radiative
lifetime, must dominate at large radii. Figure \ref{fig.misto.coma} shows, in the case of the Coma cluster,
how the mix of the two contributions alleviates
the problem of the radial behavior of $B$, even if the contribution of the $SRE$
within the thermal core radius is assumed to be as large as that of the $PRE$.

The GR production rate in the same conditions described above
immediately yields the expected luminosity and flux in the 0.1 to 10 GeV. A comparison
with the sensitivity curves of the future missions AGILE and GLAST shows
that, out of the four clusters considered here, only the Coma cluster
could be very significantly detected by GLAST, even if $\xi_0$ were as small as 0.1.
With $\xi_0$ = 0.3, Coma could be marginally detected by AGILE, 
while A2319 and A2163 could be marginally detected by GLAST.

The proposal by Bowyer et al. (2004) that the spatially-resolved UV excess emission in the
Coma Cluster, supposedly due to IC scattering by $RE$ on the cosmic microwave
background, might originate from $SRE$, is discussed as an instructive example
of what could be done if spatially-resolved measurements of the IC emission from the radioemitting
electrons, expected in the hard X-rays, were available. The $SRE$ hypothesis
is rejected on the grounds of an estimate of $\xi$ that, at the center
of the cluster, should be at least one order of magnitude greater than unity.

Finally, the dichotomy between clusters with and without radio halos
is presented as an opportunity to test the $CR$ content in
clusters in general. This follows from the fact that clusters
without a diffuse radio emission, according to FR measurements
(Clarke, Kronberg \& B\"ohringer 2001), share
a magnetic field of similar strength with
those endowed with a radio halo. The results show that, if $\xi_0$ were
close to the reference value adopted in this paper, the
$SRE$ would give rise to a measurable radio emission. Therefore,
any effort to improve the limits on diffuse radio emission from
clusters in general, and the estimate of their magnetic field
through FR measurements, will help to constrain their $CR$ content
before the hard task of the future gamma ray experiments can be
accomplished.

\acknowledgements{The authors thank G. Giovannini for early discussions of the subject and the referee
for constructive comments. S.C. is supported by PRIN-MIUR under contract No.2004027755$\_$003.}

\end{document}